\documentclass[aps,twocolumn,prb,showpacs,superscriptaddress]{revtex4}
\usepackage{graphics,bm}
\usepackage{amssymb}
\usepackage{epsfig}
\usepackage{epsf}

\def\Im{\rm{Im}}
\def\be{\begin{equation}} \def\ee{\end{equation}}
\def\bea{\begin{eqnarray}} \def\eea{\end{eqnarray}}

\def\nn{\nonumber}

\begin{document}

\title{Non-Fermi Liquid due to Orbital Fluctuations in Iron Pnictide Superconductors}

\author{Wei-Cheng Lee}
\email{leewc@illinois.edu}

\author{Philip W. Phillips}
\email{dimer@vfemmes.physics.illinois.edu}
\affiliation{Department of Physics, University of Illinois, 1110 West Green Street, Urbana, Illinois 61801, USA}

\date{\today}

\begin{abstract}
We study the influence of quantum fluctuations on the electron self energy 
in the normal state of iron-pnictide superconductors using a five orbital 
tight binding model
with generalized Hubbard on-site interactions. Within a one-loop 
treatment, we find that  an overdamped collective mode develops at low 
frequency in channels associated with quasi one-dimensional $d_{xz}$ and $d_{yz}$ 
bands.
When the critical point for the $C_4$ symmetry-broken phase (structural
phase transition) is approached, the overdamped collective modes  soften, 
and acquire increased spectral weight, resulting in
non-Fermi liquid behaviour at the Fermi surface characterized by a frequency dependence of the imaginary part of the electron self energy of the form
$\omega^\lambda$, $0<\lambda<1$.   We argue that this non-Fermi 
liquid behaviour is responsible for the recently observed zero-bias 
enhancement in the tunneling signal in
point contact spectroscopy.  A key experimental test of this 
proposal is the absence of non-Fermi liquid behaviour in the 
hole-doped materials.  Our result suggests that quantum criticality plays 
an important role
in understanding the normal state properties of iron-pnictide 
superconductors.
\end{abstract}
\pacs{}

\maketitle

\section {Introduction}
Whether or not the iron pnictide superconductors are strongly correlated 
materials is hotly debated.
Certainly a clean association of non-Fermi liquid behaviour, either 
experimentally or theoretically, with any part of the phase diagram would 
suffice to settle this debate.   While parallels with the cuprates are 
suggestive\cite{haule2008,craco2008,vildosola2008,yang2009,yin2010}, they 
have not resulted in a decisive answer to this problem.
  In fact, to our knowledge, the possibility of non-Fermi liquid behaviour 
other than Mott physics\cite{si2011} has not been discussed to date.

A feature common to the parent and underdoped compounds of the iron pnictide 
superconductors is the structural phase transition (SPT) from tetragonal 
to orthorhombic
symmetry occuring around 150K\cite{nomura2008}. For most members of the 
1111 and 122 families, in-plane anisotropy in the resistivity commences near the structural
transition, and stripe-like antiferromagnetism develops
if the temperature is further lowered. 
The quasiparticle interference in STM\cite{chuang2010} also showed aniostropic electronic states 
at low temperature. Despite the controversy as to whether 
the SPT is induced by magnetic fluctuations as a result of the onset of stripe-like 
antiferromagnetism\cite{yildirim2008,sushko2008,xu2008,chen_f2008,fernandes2010} 
or if
orbital ordering in quasi-1D $d_{xz}$ and $d_{yz}$ 
bands\cite{lv2009,kruger2009,lee_cc2009,chen_cc2010} is the efficient 
cause, the phase below the SPT breaks
$C_4$ symmetry, and quantum fluctuations associated with this phase are 
nematic in character.  A recent measurement of photoexcited
quasiparticle relaxation\cite{stoj2011} reveals the existence of strong 
nematic fluctuations up to 200K, well above the SPT temperature.
Moreover, in electron-doped Ba(Fe$_{1-x}$Co$_x$)$_2$As$_2$ (Ba122), an unexpected 
enhancement of the zero-bias signal\cite{arham2011,arham2012} in the tunneling 
data measured in point-contact spectroscopy
has been observed at an onset temperature higher than $T_{\rm SPT}$. 
The excess conductance appears at temperatures around 175K, increasing in magnitude through 
the structural, antiferromagnetic, and, in materials exhibiting superconductivity, the 
superconducting transitions. It is not seen in overdoped Ba122.
It is important to study how these strong orbital (nematic) fluctuations affect the 
physical properties in both the normal and orthorhombic states of the iron 
pnictide superconductors.

In this paper, we develop a microscopic theory for the orbital 
fluctuations and show that they give rise to non-Fermi liquid behaviour. 
In particular, we find a branch of overdamped collective modes in the
scattering channels associated with quasi-1D $d_{xz}$ and $d_{yz}$ bands 
in the normal state at a temperature higher than $T_{\rm SPT}$. In the vicinity 
of the
SPT critical point, these overdamped collective modes dominate the
low-energy
 physics, resulting in a strong modification of the electron self 
energy and a
breakdown of Fermi-liquid theory even in the symmetric normal state.

\begin{figure}
\includegraphics{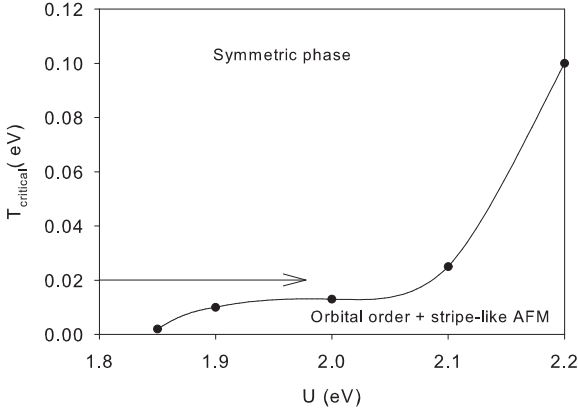}
\caption{\label{fig:mf-phase} Mean-field phase diagram of the model
  used in this paper. The arrow indicates the direction the critical
  region is approached in this paper.  At $T=0$, the orbital ordering
  transition is driven entirely by changing the strength of the
  interaction.}
\end{figure}

\section{RPA theory for electron self energy}
Since the random phase
approximation (RPA) represents a perturbative treatment of the
fluctuations,  it represents a zeroth-order theory.  If deviations
from Fermi liquid theory are found at this level of theory, then a Fermi liquid description is most likely invalid.  
This line of reasoning has been used previously in the continuum limit to establish the existence of the nematic phase \cite{lawler2006}.   
Thus far, a similar analysis appears to be lacking for a multi-orbital model.   We use this approach here to show that at the RPA level, the fluctuations are inherently non-Fermi liquid in nature.   Our starting point 
is the Hamiltonian $H=H_t + H_I$, where $H_t=\sum_{\vec{k}} H_k$ is the 
five-band tight binding model proposed in Ref. [\onlinecite{graser2009}] 
which
can reproduce correctly the Fermi surfaces of hole $\alpha_1,\alpha_2$ and 
electron $\beta_1,\beta_2$ pockets in the unfolded Brillouin zone.
The interaction terms are given by $H_I$
\bea
H_I &=& \sum_{ia} U n_{ia\uparrow}n_{ia\downarrow} + 
\sum_{i,b>a}(U'-\frac{J}{2}) n_{ia} n_{ib} - \sum_{i,b>a}2J\vec{S}_{ia}\cdot 
\vec{S}_{ib}\nn\\
&+& J'\big(p_{ia}^\dagger p_{ib} + h.c.\big),
\label{hamI}
\eea
where $U'=U-2J$, $J$ the Hund coupling, $J'=J$, and $a$ ($b$) refers to the 
orbital index, $1= xz$, $2= yz$, $3= xy$,
$4= x^2-y^2$, and $5= 3z^2-r^2$. 
$n_{ia}=\sum_{\sigma}c^\dagger_{ia\sigma} c_{ia\sigma}$ is the number operator on site $i$ in orbital $a$, and 
$p_{ia}=c_{ia\downarrow}c_{ia\uparrow}$.
Energies are measured in {\it eV}, in
line with the units used in recent tight-binding models\cite{graser2009}.
$J=0.2U$ throughout this paper.
This model has been shown to have stripe-like antiferromagnetism
together with orbital ordering in a previous study.\cite{lv2011-1,lv2011-2}
We introduce a unitary transformation $\hat{U}_{\vec{k}}$ such that
$\big(\hat{U}_{\vec{k}}\big)^\dagger \hat{H}_{\vec{k}} \hat{U}_{\vec{k}} = 
{\rm diag.}\big[E_{\vec{k},1},\cdots,E_{\vec{k},5}\big]$,
and it is straightforward to obtain the non-interacting response 
functions,
\bea
&&\chi^{(0)}_{ab;cd}(\vec{q},i\omega_n)\nn\\
&=&-\frac{1}{N}\sum_{\vec{k}}\sum_{l,m}
\big(\hat{U}_{\vec{k}+\vec{q}}\big)_{a,l}\big(\hat{U}_{\vec{k}+\vec{q}}\big)^*_{c,l}
\big(\hat{U}_{\vec{k}}\big)_{d,m}\big(\hat{U}_{\vec{k}}\big)^*_{b,m}\nn\\
&\times&\frac{n_F(E_{\vec{k}+\vec{q},l}) - 
n_F(E_{\vec{k},m})}{E_{\vec{k}+\vec{q},l} - E_{\vec{k},m} - i\omega_n},
\eea
in the symmetric normal phase, where 
$\chi^{(0)}_{ab;cd}(\vec{q},i\omega_n)$ is a $25\times 25$ matrix.  The 
convention on the indices is that in Ref.[\onlinecite{graser2009}].
Adopting the interaction kernels for the spin-spin ($\hat{V}^s$) and 
density-density ($\hat{V}^c$) fluctuations derived in 
Ref.[\onlinecite{kemper2010}], we
obtain the electron self energy at the random-phase approximation level,
\be
\Sigma^{orbital}_{ab}(\vec{k},ip_n) = \frac{1}{\beta 
N}\sum_{\vec{q}}\sum_{ik_m} \hat{\Gamma}_{a i;j 
d}(\vec{q},ip_n-ik_m)\hat{G}^0_{ij}(\vec{k}-\vec{q},ik_m)
\label{self-1},
\ee
where $ik_m$ and $ip_n$ are Matsubara frequencies for fermions, 
$\hat{G}^0(\vec{k},ip_n) = \big[ip_n - \hat{H}_t + \mu\big]^{-1}$ is the
bare Green function, and $\hat{\Gamma}_{ab;cd}(\vec{q},ip_n-ik_m)$ is
the effective interaction,
\bea
\hat{\Gamma}(\vec{q},ip_n-ik_m) &=& \frac{1}{2}\big\{ 
3\big[1-\hat{V}^s\hat{\chi}_0(\vec{q},ip_n-ik_m)\big]^{-1}\hat{V}^s\nn\\
&-&\big[1+\hat{V}^c\hat{\chi}_0(\vec{q},ip_n-ik_m)\big]^{-1}\hat{V}^c\big\}
\label{eq:gamma}
\eea
  within one-loop.  

\section{Critical behaviour above the structural phase transition}
Since our focus is on the critical region above the structural phase transition, throughout this paper the temperature is set to 
$k_B T=0.02$eV at which the system is in a normal state without any symmetry breaking. 
The case of ordered states will be discussed in the next section.

\begin{figure}
\includegraphics{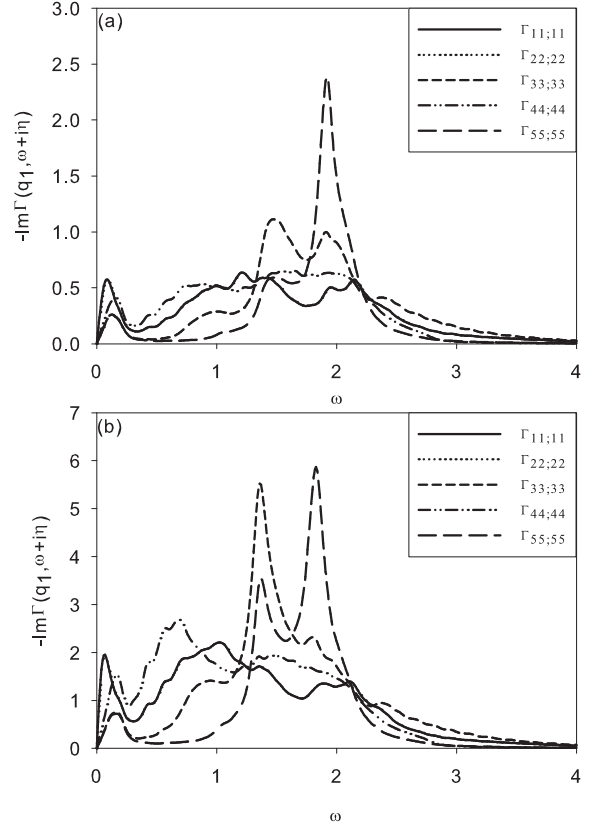}
\caption{\label{fig:spectral-all} Spectral functions for effective
interactions in intra-orbital channels ($\Gamma_{aa;aa}(\vec{q},\omega)$)
at $\vec{q}_1= (0.04\pi,0.04\pi)$ for
(a) U=1.3 and (b) U=2.0.
$\Gamma_{11;11}(\vec{q},\omega)=\Gamma_{22;22}(\vec{q},\omega)$ for
$\vec{q}$ along the diagonal direction as expected, and an overdamped
collective mode
appearing at low energy in
$\Gamma_{11;11}(\vec{q},\omega)(\Gamma_{22;22}(\vec{q},\omega))$ can be
observed.}
\end{figure}

\begin{figure}
\includegraphics{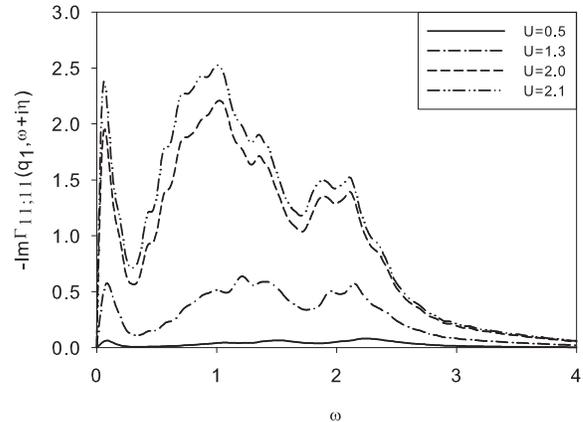}
\caption{\label{fig:spectral-xz} Spectral functions of effective
interactions $\Gamma_{11;11}(\vec{q}_1,\omega)$ at $\vec{q}_1=
(0.04\pi,0.04\pi)$ for
different $U$. The overdamped collective mode at low energy acquires
increased spectral weight as the critical point ($U_c\approx 2.1 eV$ at the temperature we are considering $k_B T=0.02 eV$) is approached.}
\end{figure}

We start by studying the mean-field phase diagram of our model using the formalism outlined in Ref. [\onlinecite{lv2011-1}], as shown in Fig. \ref{fig:mf-phase}.
Then we can fix the temperature and change $U$ to investigate how the self-energy changes as the critical region is approached.
Consider first the spectral functions for the effective interactions 
$-{\rm Im}\hat{\Gamma}(\vec{q},\omega+i\eta)$ displayed in
Fig. \ref{fig:spectral-all}.   As is evident, the spectral functions for 
the intra-orbital effective interactions $-{\rm 
Im}\Gamma_{aa;aa}(\vec{q}_1,\omega)$
dominate the electron self energy.
Also of interest are the spectral functions for momenta $\vec{q}$ along 
the diagonal direction so that 
$\Gamma_{11;11}(\vec{q},\omega)=\Gamma_{22;22}(\vec{q},\omega)$.
It can be seen in Fig. \ref{fig:spectral-all} that the spectral functions 
at low frequency are dominated by an overdamped collective mode in 
$\Gamma_{11;11}$
($\Gamma_{22;22}$).
When $U$ is tuned to approach the critical point ($U_c\approx 2.1 eV$ at this 
temperature), this mode gains more spectral weight and moves to even lower 
energy as shown in Fig. \ref{fig:spectral-xz}.
Note that although the shift of the spectral weight upon approaching the critical point is shown to follow the typical behaviour for an overdamped 
continuum of collective excitations, our calculations do not have the accuracy to demonstrate the shift of the peak position toward zero frequency as expected.
We attribute this to the fact that the model considered here generally has both orbital ordering and stripe-AFM phases occuring together. 
In this case, it is hard to distinguish whether the transition is first or second order as discussed in Ref. [\onlinecite{preempty}], which is beyond the accuracy of our 
calculation on a finite-size square lattice.

These overdamped modes, emergent at low frequency and small $\vec{q}$, 
resemble the collective modes observed in the {\it quadrupole density 
spectral
function}\cite{oganesyan2001,yamase2004,lawler2006,kao2007},  that is, the spectral 
function related to the interactions in the $d$-wave channel in a
quantum nematic Fermi fluid. As shown in Ref. 
[\onlinecite{lee2009nematic}], in a system containing
quasi-1D $d_{xz}$ and $d_{yz}$ bands, hybridization enhances 
significantly the strength of the interaction in the $d$-wave channel.
As a result, the nematic order in such multiorbital systems is completely 
{\bf equivalent} to orbital ordering in quasi-1D bands, and the spectral functions due to quantum 
fluctuations associated with the quasi-1D bands naturally acquire the
same properties of the quadrupole density
spectral function discussed in the context of the quantum nematic fluid mentioned above.
It has been shown\cite{oganesyan2001,lawler2006,kao2007} that these 
overdamped collective modes could lead to a non-Fermi liquid near the 
critical region and also in
the nematic phase. The reason is that in the vicinity of the nematic 
critical point, these overdamped collective modes become soft.
Electrons scatter strongly with these soft overdamped collective modes, 
which modify the electron self energy away from the Fermi liquid 
behaviour in the vicinity
of the nematic critical point.

It is intriguing to check whether the same physics discussed above occurs 
in the iron-pnictide superconductors since the SPT signals a transition
from thei symmetric normal phase to a state which breaks $C_4$ symmetry.  We 
performed a numerical evaluation of Eq. \ref{self-1}. To compute the self 
energy of the retarded Green function of a single-particle state on the 
Fermi surface,
we need to do one more transformation and also an analytical continuation 
on Eq. \ref{self-1} to obtain
\be
\Sigma^{band}_{\alpha\alpha}(\vec{k}_F,\omega+i\eta) = 
\big(\hat{U}_{\vec{k}_F}\hat{\Sigma}^{orbital}(\vec{k}_F,\omega+i\eta)\hat{U}^\dagger_{\vec{k}_F}\big)_{\alpha\alpha},
\ee
which is the self energy of the electron with momentum $\vec{k}_F$ on the 
Fermi surface sheet $\alpha$.

\begin{figure}
\includegraphics{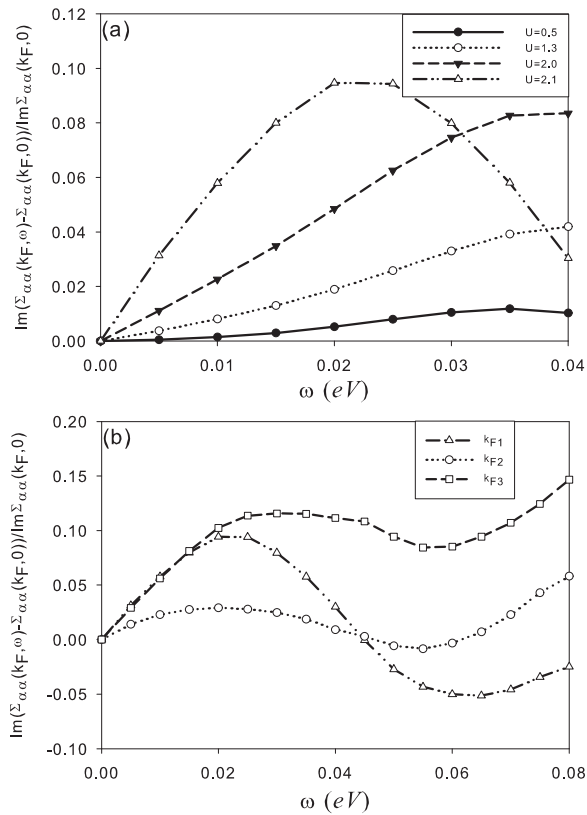}
\caption{\label{fig:self} (a)Normalized self-energy of electron with 
momentum $\vec{k}_F=(0.12\pi,0.12\pi)$ on the hole Fermi pocket $\alpha_1$ 
for
different $U$. A crossover from Fermi liquid ($\lambda=2$) to non-Fermi 
liquid ($\lambda\leq 1$) can be seen as $U$ increases from $U=0.5$ to 
critical point $U=2.1 eV$.
(b) Normalized self-energy of electron in the critical region (U=2.1 eV) with 
momenta $\vec{k}_{F1}=(0.12\pi,0.12\pi)$, $\vec{k}_{F2}=(0.2\pi,0)$
on hole Fermi pocket $\alpha_1$, and $\vec{k}_{F3}=(0.88\pi,0.16\pi)$ on 
electron Fermi pocket $\beta_1$.}
\end{figure}

Because we study the normal state at finite temperature,
$\Sigma^{band}_{\alpha\alpha}(\vec{k}_F,\omega+i\eta)$ contains
contributions from both
thermal and quantum fluctuations which can not be separated in RPA-type calculations\cite{kao2007}. 
Nevertheless, it is generally expected that the contribution from quantum critical fluctuations
should be expanded in powers of $\omega/T$ and the thermal fluctuations should be most dominant at $\omega=0$.
In order to see the frequency dependence due to the quantum critical fluctuations more clearly, we 
plot in Fig. \ref{fig:self} the normalized imaginary part of the self energy defined as
\be
{\rm Im}\Sigma^{nor}_{\alpha\alpha}(\vec{k}_F,\omega)\equiv
\frac{\big({\Im} \Sigma^{band}_{\alpha\alpha}(\vec{k}_F,\omega) - {\Im} 
\Sigma^{band}_{\alpha\alpha}(\vec{k}_F,0) \big)}{{\Im} 
\Sigma^{band}_{\alpha\alpha}(\vec{k}_F,0)}.
\ee
Generally, it is expected that at finite temperature the self-energy
is analytical for $\hbar \omega << k_B T$, which gives the $\omega^2$ term.
We find the crossover from a Fermi liquid with $\omega^2$ at small frequency to a non-Fermi liquid 
in which the $\omega^\lambda$ term with $\lambda\leq 1$ dominates over the $\omega^2$ term as $U$ is increased to approach 
the critical point. Similar results have been seen in Ref. [\onlinecite{kao2007}].
In the critical region,  non-Fermi liquid behaviour exists on a 
large part of the Fermi surface with strong angular dependence of ${\rm 
Im}\Sigma$ as
expected due to the critical fluctuations near orbital ordering (now termed 
nematicity)\cite{oganesyan2001}.
As the temperature is lowered, the critical point $U_c$ shifts to a lower 
value but the non-Fermi liquid behaviour remains robust near the critical 
point.
This strongly suggest that this non-Fermi liquid behaviour should  be 
observable in iron-pnictide superconductors at a temperature above the SPT.

One subtle point is that the effective interaction $\Gamma$ in
Eq. \ref{eq:gamma} contains contributions from both the charge and spin channels.
While for large momentum (e.g. $(\pi,0)$ or $(0,\pi)$) there is no doubt that the contribution from spin channels is dominant as seen from
previous calculations\cite{graser2009,kemper2010} and also in our
calculations, for  small momentum, which we focus on here, the contributions
from both charge and spin channels become comparable. As a result, we can not completely rule out the effects from the spin channels.
However, since it has been shown\cite{preempty} that the spin nematicity could also induce  orbital ordering, the critical collective modes associated
with the orbital ordering discussed above will be still present in
that case, despite the fact that the spectral weight might be reduced due to the coupling to
the collective modes associated with spin nematic order.
As a result, the non-Fermi liquid behaviour discussed above will be most prominent if the contribution from charge channels in Eq. \ref{eq:gamma} is dominant.
We find, in general, that the charge channels become much stronger for $U\approx U'$, which is consistent with previous studies of orbital ordering in Sr$_3$Ru$_2$O$_7$
\cite{leewc2009,raghu2009}.
Moreover, it has been shown that the inclusion of electron-phonon coupling can also enhance the instability in the charge channels\cite{onari2010}.
Since it is well-known that physical properties of iron-based superconductors could vary significantly for different families due to difference in details, we
expect that the non-Fermi liquid behaviour discussed above might not be visible in some families of the iron-based superconductors.
This is actually what is seen in  point contact spectroscopy and other experiments which will be discussed in Sec. V.

\section{Non-Fermi liquid behaviour below the structural phase transition}
In this section, we discuss the fate of the non-Fermi liquid behaviour in the C$_4$ symmetry broken phase. 
For the nematic phase in a continuous model, these overdamped
collective modes induced by the $d$-wave interaction evolve into Goldstone modes but remain overdamped and
dominate the low energy physics.
Consequently, the non-Fermi liquid persists in the nematic phase.
In the multiorbital model studied here, the situation is complicated by the fact that since the continuous rotational symmetry is absent in a lattice model, 
there are no gapless Goldstone modes in general.
Nevertheless, these overdamped collective modes remain existing with a gap $\Delta$ due to the breaking of the discrete symmetry from $C_4$ to $C_2$.
Consequently, the non-Fermi liquid behaviour will be present as long as the temperature energy scale $k_B T$ is larger than $\Delta$. 
Note that $\Delta$ is a gap in the density-density correlation function not in the single particle spectrum, since the orbital order (like the nematic order) 
does not gap out the Fermi surfaces.
As a result, it is expected that the non-Fermi liquid will persist for a while as the temperature is lowered below $T_{\rm SPT}$ and then 
gradually disappear at very low temperature where the orbital order is strong.
This is analogous to the case of the ferromagnetic quantum critical point with a magnetic field where the critical fluctuations are gapped by the Zeeman energy\cite{coleman}.

\section{Experimental Consequences} 
A direct cconsequence of non-Fermi 
liquid behaviour is the temperature dependence of the resistivity.
It has been pointed out\cite{ying2011} from studies on various 
iron-pnictide superconductors that a strong deviation from the Fermi
liquid $T^2$ behaviour of the resistivity 
above the
SPT temperature will occur if a large anisotropy in the in-plane resistivity 
exists below the SPT temperature.
Since the anisotropy in resistivity is intimately related to orbital ordering, this observation provides direct evidence for our claim that non-Fermi liquid
behaviour is due to orbital fluctuations.
An independent calculation including electron-phonon coupling to enhance the effect of orbital fluctuations 
by Onari and Kontani\cite{onari2010} also showed 
unusual temperature dependence of the resistivity above the structural transition temperature.

What about the zero-bias anomaly seen in point-contact tunneling
experiments\cite{arham2011,arham2012} on electron-doped 
Ba(Fe$_{1-x}$Co$_x$)$_2$As$_2$?
Intriguingly, this zero-bias enhancement starts to appear at temperatures 
roughly 30K higher than $T_{\rm SPT}$
and remains robust well below $T_{\rm SPT}$.
This observation is also consistent with our theory.
It has been shown by Lawler {\it et. al.}\cite{lawler2006} that the single 
particle density of states has the form of
\be
N^*(\omega) = N^*(0) + B \omega^{2/3}\ln\omega + \cdots
\ee
in the nematic critical region and also in the nematic phase. In fact, 
$N^*(\omega)$ obtains extra contributions due to the non-Fermi liquid
self-energy, giving rise to
a cusp at zero frequency and a subsequent decrease as the frequency 
increases. This provides a direct explanation for the zero-bias 
enhancement observed in point contact
spectroscopy since the conductance $dI/dV$ roughly measures  the single-particle density of states for small frequency.
Moreover, since the form of the single-particle density of states is the 
same up to some mild modifications in the vicinity of the critical point 
and also
in the $C_4$ symmetry broken phase, the zero-bias enhancement should have 
a smooth crossover
as $T_{\rm SPT}$ is crossed, which in fact has been noticed in 
quantum point-contact measurements\cite{note}.
We predict that for hole-underdoped
Ba$_{1-x}$K$_x$Fe$_2$As$_2$ which does not have an in-plane resistivity 
anisotropy\cite{ying2011}, the zero-bias enhancement should be either nonexistent or much weaker than that in electron-doped 
Ba(Fe$_{1-x}$Co$_x$)$_2$As$_2$.

\section{Conclusion} 
In this paper we have presented a theory of non-Fermi 
liquid behaviour in a five-band model with generalized Hubbard on-site interactions 
for
iron pnictide superconductors. At the level of the random-phase
approximation, we  found a branch of overdamped collective modes emergent 
at low frequency in channels associated with quasi-1D $d_{xz}$ and 
$d_{yz}$ bands, and
we have shown that these modes become dominant at low energies near the 
critical point for the $C_4$ symmetry-broken phase, leading to non-Fermi 
liquid behaviour.
Our theory indicates that quantum criticality through the evolution of
a non-Fermi liquid phase plays an
important role in understanding the normal-state properties of iron-pnictide superconductors.

\section{Acknowledgment} 
We would like to thank H.Z. Arham, Piers Coleman, Eduardo Fradkin, Laura H. 
Greene, Jiangping Hu, Anthony J. Leggett, Weicheng Lv, W.K. Park, and Qimiao Si for helpful 
discussions. We are particularly grateful to Andrey Chubukov for many of his very critical suggestions and comments on this paper.
This work is supported by the Center for Emergent 
Superconductivity, a DOE Energy Frontier Research Center, Grant 
No.~DE-AC0298CH1088.
W.-C. L. would like to thank Aspen Center for Physics supported by NSF 
Grant No. 1066293 for the hospitality while this work was initiated.


\begin{thebibliography}{26}
\expandafter\ifx\csname natexlab\endcsname\relax\def\natexlab#1{#1}\fi
\expandafter\ifx\csname bibnamefont\endcsname\relax
   \def\bibnamefont#1{#1}\fi
\expandafter\ifx\csname bibfnamefont\endcsname\relax
   \def\bibfnamefont#1{#1}\fi
\expandafter\ifx\csname citenamefont\endcsname\relax
   \def\citenamefont#1{#1}\fi
\expandafter\ifx\csname url\endcsname\relax
   \def\url#1{\texttt{#1}}\fi
\expandafter\ifx\csname urlprefix\endcsname\relax\def\urlprefix{URL }\fi
\providecommand{\bibinfo}[2]{#2}
\providecommand{\eprint}[2][]{\url{#2}}

\bibitem[{\citenamefont{Haule et~al.}(2008)\citenamefont{Haule, Shim, and
   Kotliar}}]{haule2008}
\bibinfo{author}{\bibfnamefont{K.}~\bibnamefont{Haule}},
   \bibinfo{author}{\bibfnamefont{J.~H.} \bibnamefont{Shim}}, 
\bibnamefont{and}
   \bibinfo{author}{\bibfnamefont{G.}~\bibnamefont{Kotliar}},
   \bibinfo{journal}{Phys. Rev. Lett.} \textbf{\bibinfo{volume}{100}},
   \bibinfo{pages}{226402} (\bibinfo{year}{2008}).

\bibitem[{\citenamefont{Craco et~al.}(2008)\citenamefont{Craco, Laad, 
Leoni,
   and Rosner}}]{craco2008}
\bibinfo{author}{\bibfnamefont{L.}~\bibnamefont{Craco}},
   \bibinfo{author}{\bibfnamefont{M.~S.} \bibnamefont{Laad}},
   \bibinfo{author}{\bibfnamefont{S.}~\bibnamefont{Leoni}}, 
\bibnamefont{and}
   \bibinfo{author}{\bibfnamefont{H.}~\bibnamefont{Rosner}},
   \bibinfo{journal}{Phys. Rev. B} \textbf{\bibinfo{volume}{78}},
   \bibinfo{pages}{134511} (\bibinfo{year}{2008}).

\bibitem[{\citenamefont{Vildosola et~al.}(2008)\citenamefont{Vildosola,
   Pourovskii, Arita, Biermann, and Georges}}]{vildosola2008}
\bibinfo{author}{\bibfnamefont{V.}~\bibnamefont{Vildosola}},
   \bibinfo{author}{\bibfnamefont{L.}~\bibnamefont{Pourovskii}},
   \bibinfo{author}{\bibfnamefont{R.}~\bibnamefont{Arita}},
   \bibinfo{author}{\bibfnamefont{S.}~\bibnamefont{Biermann}}, 
\bibnamefont{and}
   \bibinfo{author}{\bibfnamefont{A.}~\bibnamefont{Georges}},
   \bibinfo{journal}{Phys. Rev. B} \textbf{\bibinfo{volume}{78}},
   \bibinfo{pages}{064518} (\bibinfo{year}{2008}).

\bibitem[{\citenamefont{Yang et~al.}(2009)\citenamefont{Yang, Sorini, 
Chen,
   Moritz, Lee, Vernay, Olalde-Velasco, Denlinger, Delley, Chu
   et~al.}}]{yang2009}
\bibinfo{author}{\bibfnamefont{W.~L.} \bibnamefont{Yang}},
   \bibinfo{author}{\bibfnamefont{A.~P.} \bibnamefont{Sorini}},
   \bibinfo{author}{\bibfnamefont{C.-C.} \bibnamefont{Chen}},
   \bibinfo{author}{\bibfnamefont{B.}~\bibnamefont{Moritz}},
   \bibinfo{author}{\bibfnamefont{W.-S.} \bibnamefont{Lee}},
   \bibinfo{author}{\bibfnamefont{F.}~\bibnamefont{Vernay}},
   \bibinfo{author}{\bibfnamefont{P.}~\bibnamefont{Olalde-Velasco}},
   \bibinfo{author}{\bibfnamefont{J.~D.} \bibnamefont{Denlinger}},
   \bibinfo{author}{\bibfnamefont{B.}~\bibnamefont{Delley}},
   \bibinfo{author}{\bibfnamefont{J.-H.} \bibnamefont{Chu}},
   \bibnamefont{et~al.}, \bibinfo{journal}{Phys. Rev. B}
   \textbf{\bibinfo{volume}{80}}, \bibinfo{pages}{014508}
   (\bibinfo{year}{2009}).

\bibitem[{\citenamefont{Yin et~al.}(2010)\citenamefont{Yin, Lee, and
   Ku}}]{yin2010}
\bibinfo{author}{\bibfnamefont{W.-G.} \bibnamefont{Yin}},
   \bibinfo{author}{\bibfnamefont{C.-C.} \bibnamefont{Lee}}, 
\bibnamefont{and}
   \bibinfo{author}{\bibfnamefont{W.}~\bibnamefont{Ku}}, 
\bibinfo{journal}{Phys.
   Rev. Lett.} \textbf{\bibinfo{volume}{105}}, \bibinfo{pages}{107004}
   (\bibinfo{year}{2010}).

\bibitem{si2011} Elihu Abrahams and Qimiao Si, J. Phys.: Condens. Matt. {\bf 23} 223201 (2011).

\bibitem[{\citenamefont{Nomura et~al.}(2008)\citenamefont{Nomura, Kim,
   Kamihara, Hirano, Sushko, Kato, Takata, Shluger, and 
Hosono}}]{nomura2008}
\bibinfo{author}{\bibfnamefont{T.}~\bibnamefont{Nomura}},
   \bibinfo{author}{\bibfnamefont{S.}~\bibnamefont{Kim}},
   \bibinfo{author}{\bibfnamefont{Y.}~\bibnamefont{Kamihara}},
   \bibinfo{author}{\bibfnamefont{M.}~\bibnamefont{Hirano}},
   \bibinfo{author}{\bibfnamefont{P.~V.} \bibnamefont{Sushko}},
   \bibinfo{author}{\bibfnamefont{K.}~\bibnamefont{Kato}},
   \bibinfo{author}{\bibfnamefont{M.}~\bibnamefont{Takata}},
   \bibinfo{author}{\bibfnamefont{A.~L.} \bibnamefont{Shluger}},
   \bibnamefont{and} 
\bibinfo{author}{\bibfnamefont{H.}~\bibnamefont{Hosono}},
   \bibinfo{journal}{Supercond. Sci. Technol.} 
\textbf{\bibinfo{volume}{21}},
   \bibinfo{pages}{125028} (\bibinfo{year}{2008}).

\bibitem[{\citenamefont{Chuang et~al.}(2010)\citenamefont{Chuang, Allan, 
Lee,
   Xie, Ni, Bud'ko, Boebinger, Canfield, and Davis}}]{chuang2010}
\bibinfo{author}{\bibfnamefont{T.-M.} \bibnamefont{Chuang}},
   \bibinfo{author}{\bibfnamefont{M.~P.} \bibnamefont{Allan}},
   \bibinfo{author}{\bibfnamefont{J.}~\bibnamefont{Lee}},
   \bibinfo{author}{\bibfnamefont{Y.}~\bibnamefont{Xie}},
   \bibinfo{author}{\bibfnamefont{N.}~\bibnamefont{Ni}},
   \bibinfo{author}{\bibfnamefont{S.~L.} \bibnamefont{Bud'ko}},
   \bibinfo{author}{\bibfnamefont{G.~S.} \bibnamefont{Boebinger}},
   \bibinfo{author}{\bibfnamefont{P.~C.} \bibnamefont{Canfield}},
   \bibnamefont{and} \bibinfo{author}{\bibfnamefont{J.~C.} 
\bibnamefont{Davis}},
   \bibinfo{journal}{Science} \textbf{\bibinfo{volume}{327}},
   \bibinfo{pages}{181} (\bibinfo{year}{2010}).

\bibitem[{\citenamefont{Yildirim}(2008)}]{yildirim2008}
\bibinfo{author}{\bibfnamefont{T.}~\bibnamefont{Yildirim}},
   \bibinfo{journal}{Phys. Rev. Lett.} \textbf{\bibinfo{volume}{101}},
   \bibinfo{pages}{057010} (\bibinfo{year}{2008}).

\bibitem[{\citenamefont{Sushko et~al.}(2008)\citenamefont{Sushko, Shluger,
   Hirano, and Hosono}}]{sushko2008}
\bibinfo{author}{\bibfnamefont{P.~V.} \bibnamefont{Sushko}},
   \bibinfo{author}{\bibfnamefont{A.~L.} \bibnamefont{Shluger}},
   \bibinfo{author}{\bibfnamefont{M.}~\bibnamefont{Hirano}}, 
\bibnamefont{and}
   \bibinfo{author}{\bibfnamefont{H.}~\bibnamefont{Hosono}},
   \bibinfo{journal}{Phys. Rev. B} \textbf{\bibinfo{volume}{78}},
   \bibinfo{pages}{172508} (\bibinfo{year}{2008}).

\bibitem[{\citenamefont{Xu et~al.}(2008)\citenamefont{Xu, M\"uller, and
   Sachdev}}]{xu2008}
\bibinfo{author}{\bibfnamefont{C.}~\bibnamefont{Xu}},
   \bibinfo{author}{\bibfnamefont{M.}~\bibnamefont{M\"uller}}, 
\bibnamefont{and}
   \bibinfo{author}{\bibfnamefont{S.}~\bibnamefont{Sachdev}},
   \bibinfo{journal}{Phys. Rev. B} \textbf{\bibinfo{volume}{78}},
   \bibinfo{pages}{020501} (\bibinfo{year}{2008}).

\bibitem[{\citenamefont{Fang et~al.}(2008)\citenamefont{Fang, Yao, Tsai, 
Hu,
   and Kivelson}}]{chen_f2008}
\bibinfo{author}{\bibfnamefont{C.}~\bibnamefont{Fang}},
   \bibinfo{author}{\bibfnamefont{H.}~\bibnamefont{Yao}},
   \bibinfo{author}{\bibfnamefont{W.-F.} \bibnamefont{Tsai}},
   \bibinfo{author}{\bibfnamefont{J.}~\bibnamefont{Hu}}, \bibnamefont{and}
   \bibinfo{author}{\bibfnamefont{S.~A.} \bibnamefont{Kivelson}},
   \bibinfo{journal}{Phys. Rev. B} \textbf{\bibinfo{volume}{77}},
   \bibinfo{pages}{224509} (\bibinfo{year}{2008}).

\bibitem[{\citenamefont{Fernandes et~al.}(2010)\citenamefont{Fernandes,
   VanBebber, Bhattacharya, Chandra, Keppens, Mandrus, McGuire, Sales, 
Sefat,
   and Schmalian}}]{fernandes2010}
\bibinfo{author}{\bibfnamefont{R.~M.} \bibnamefont{Fernandes}},
   \bibinfo{author}{\bibfnamefont{L.~H.} \bibnamefont{VanBebber}},
   \bibinfo{author}{\bibfnamefont{S.}~\bibnamefont{Bhattacharya}},
   \bibinfo{author}{\bibfnamefont{P.}~\bibnamefont{Chandra}},
   \bibinfo{author}{\bibfnamefont{V.}~\bibnamefont{Keppens}},
   \bibinfo{author}{\bibfnamefont{D.}~\bibnamefont{Mandrus}},
   \bibinfo{author}{\bibfnamefont{M.~A.} \bibnamefont{McGuire}},
   \bibinfo{author}{\bibfnamefont{B.~C.} \bibnamefont{Sales}},
   \bibinfo{author}{\bibfnamefont{A.~S.} \bibnamefont{Sefat}}, 
\bibnamefont{and}
   \bibinfo{author}{\bibfnamefont{J.}~\bibnamefont{Schmalian}},
   \bibinfo{journal}{Phys. Rev. Lett.} \textbf{\bibinfo{volume}{105}},
   \bibinfo{pages}{157003} (\bibinfo{year}{2010}).

\bibitem[{\citenamefont{Lv et~al.}(2009)\citenamefont{Lv, Wu, and
   Phillips}}]{lv2009}
\bibinfo{author}{\bibfnamefont{W.}~\bibnamefont{Lv}},
   \bibinfo{author}{\bibfnamefont{J.}~\bibnamefont{Wu}}, \bibnamefont{and}
   \bibinfo{author}{\bibfnamefont{P.}~\bibnamefont{Phillips}},
   \bibinfo{journal}{Phys. Rev. B} \textbf{\bibinfo{volume}{80}},
   \bibinfo{pages}{224506} (\bibinfo{year}{2009}).

\bibitem[{\citenamefont{Kr\"uger et~al.}(2009)\citenamefont{Kr\"uger, 
Kumar,
   Zaanen, and van~den Brink}}]{kruger2009}
\bibinfo{author}{\bibfnamefont{F.}~\bibnamefont{Kr\"uger}},
   \bibinfo{author}{\bibfnamefont{S.}~\bibnamefont{Kumar}},
   \bibinfo{author}{\bibfnamefont{J.}~\bibnamefont{Zaanen}}, 
\bibnamefont{and}
   \bibinfo{author}{\bibfnamefont{J.}~\bibnamefont{van~den Brink}},
   \bibinfo{journal}{Phys. Rev. B} \textbf{\bibinfo{volume}{79}},
   \bibinfo{pages}{054504} (\bibinfo{year}{2009}).

\bibitem[{\citenamefont{Lee et~al.}(2009)\citenamefont{Lee, Yin, and
   Ku}}]{lee_cc2009}
\bibinfo{author}{\bibfnamefont{C.-C.} \bibnamefont{Lee}},
   \bibinfo{author}{\bibfnamefont{W.-G.} \bibnamefont{Yin}}, 
\bibnamefont{and}
   \bibinfo{author}{\bibfnamefont{W.}~\bibnamefont{Ku}}, 
\bibinfo{journal}{Phys.
   Rev. Lett.} \textbf{\bibinfo{volume}{103}}, \bibinfo{pages}{267001}
   (\bibinfo{year}{2009}).

\bibitem[{\citenamefont{Chen et~al.}(2010)\citenamefont{Chen, Maciejko, 
Sorini,
   Moritz, Singh, and Devereaux}}]{chen_cc2010}
\bibinfo{author}{\bibfnamefont{C.-C.} \bibnamefont{Chen}},
   \bibinfo{author}{\bibfnamefont{J.}~\bibnamefont{Maciejko}},
   \bibinfo{author}{\bibfnamefont{A.~P.} \bibnamefont{Sorini}},
   \bibinfo{author}{\bibfnamefont{B.}~\bibnamefont{Moritz}},
   \bibinfo{author}{\bibfnamefont{R.~R.~P.} \bibnamefont{Singh}},
   \bibnamefont{and} \bibinfo{author}{\bibfnamefont{T.~P.}
   \bibnamefont{Devereaux}}, \bibinfo{journal}{Phys. Rev. B}
   \textbf{\bibinfo{volume}{82}}, \bibinfo{pages}{100504}
   (\bibinfo{year}{2010}).

\bibitem[{\citenamefont{{Stojchevska} 
et~al.}(2011)\citenamefont{{Stojchevska},
   {Mertelj}, {Chu}, {Fisher}, and {Mihailovic}}}]{stoj2011}
\bibinfo{author}{\bibfnamefont{L.}~\bibnamefont{{Stojchevska}}},
   \bibinfo{author}{\bibfnamefont{T.}~\bibnamefont{{Mertelj}}},
   \bibinfo{author}{\bibfnamefont{J.-H.} \bibnamefont{{Chu}}},
   \bibinfo{author}{\bibfnamefont{I.~R.} \bibnamefont{{Fisher}}},
   \bibnamefont{and}
   \bibinfo{author}{\bibfnamefont{D.}~\bibnamefont{{Mihailovic}}},
   \bibinfo{howpublished}{arXiv.org:1107.5934} (\bibinfo{year}{2011}).

\bibitem[{\citenamefont{Arham et~al.}(2011)\citenamefont{Arham, Hunt, 
Park,
   Gillett, Das, Sebastian, Xu, Wen, Lin, Li et~al.}}]{arham2011}
\bibinfo{author}{\bibfnamefont{H.~Z.} \bibnamefont{Arham}},
   \bibinfo{author}{\bibfnamefont{C.~R.} \bibnamefont{Hunt}},
   \bibinfo{author}{\bibfnamefont{W.~K.} \bibnamefont{Park}},
   \bibinfo{author}{\bibfnamefont{J.}~\bibnamefont{Gillett}},
   \bibinfo{author}{\bibfnamefont{S.~D.} \bibnamefont{Das}},
   \bibinfo{author}{\bibfnamefont{S.~E.} \bibnamefont{Sebastian}},
   \bibinfo{author}{\bibfnamefont{Z.~J.} \bibnamefont{Xu}},
   \bibinfo{author}{\bibfnamefont{J.~S.} \bibnamefont{Wen}},
   \bibinfo{author}{\bibfnamefont{Z.~W.} \bibnamefont{Lin}},
   \bibinfo{author}{\bibfnamefont{Q.}~\bibnamefont{Li}}, 
\bibnamefont{et~al.},
   \bibinfo{howpublished}{arXiv.org:1108.2749} (\bibinfo{year}{2011}).

\bibitem{arham2012} H. Z. Arham, C. R. Hunt, W. K. Park, J. Gillett, S. D. Das, S. E. Sebastian, Z. J. Xu, J. S. Wen, Z. W. Lin, Q. Li, G. Gu, A. Thaler, S. Ran, S. L. Bud'ko, P. C. Canfield, D. Y. Chung, M. G. Kanatzidis, and L. H. Greene, arXiv.org:1201.2479 (2012).

\bibitem[{\citenamefont{Lawler et~al.}(2006)\citenamefont{Lawler, Barci,
   Fern\'andez, Fradkin, and Oxman}}]{lawler2006}
\bibinfo{author}{\bibfnamefont{M.~J.} \bibnamefont{Lawler}},
   \bibinfo{author}{\bibfnamefont{D.~G.} \bibnamefont{Barci}},
   \bibinfo{author}{\bibfnamefont{V.}~\bibnamefont{Fern\'andez}},
   \bibinfo{author}{\bibfnamefont{E.}~\bibnamefont{Fradkin}},
\bibnamefont{and}
   \bibinfo{author}{\bibfnamefont{L.}~\bibnamefont{Oxman}},
   \bibinfo{journal}{Phys. Rev. B} \textbf{\bibinfo{volume}{73}},
   \bibinfo{pages}{085101} (\bibinfo{year}{2006}).

\bibitem[{\citenamefont{Graser et~al.}(2009)\citenamefont{Graser, Maier,
   Hirschfeld, and Scalapino}}]{graser2009}
\bibinfo{author}{\bibfnamefont{S.}~\bibnamefont{Graser}},
   \bibinfo{author}{\bibfnamefont{T.~A.} \bibnamefont{Maier}},
   \bibinfo{author}{\bibfnamefont{P.~J.} \bibnamefont{Hirschfeld}},
   \bibnamefont{and} \bibinfo{author}{\bibfnamefont{D.~J.}
   \bibnamefont{Scalapino}}, \bibinfo{journal}{New Journal of Physics}
   \textbf{\bibinfo{volume}{11}}, \bibinfo{pages}{025016}
   (\bibinfo{year}{2009}).

\bibitem{lv2011-1} W. Lv and P. Phillips, Phys. Rev. B 84, 174512 (2011).

\bibitem{lv2011-2} W. Lv, W.-C. Lee, and P. Phillips, Phys. Rev. B 84, 155107 (2011).

\bibitem[{\citenamefont{Kemper et~al.}(2010)\citenamefont{Kemper, Maier,
   Graser, Cheng, Hirschfeld, and Scalapino}}]{kemper2010}
\bibinfo{author}{\bibfnamefont{A.~F.} \bibnamefont{Kemper}},
   \bibinfo{author}{\bibfnamefont{T.~A.} \bibnamefont{Maier}},
   \bibinfo{author}{\bibfnamefont{S.}~\bibnamefont{Graser}},
   \bibinfo{author}{\bibfnamefont{H.-P.} \bibnamefont{Cheng}},
   \bibinfo{author}{\bibfnamefont{P.~J.} \bibnamefont{Hirschfeld}},
   \bibnamefont{and} \bibinfo{author}{\bibfnamefont{D.~J.}
   \bibnamefont{Scalapino}}, \bibinfo{journal}{New Journal of Physics}
   \textbf{\bibinfo{volume}{12}}, \bibinfo{pages}{073030}
   (\bibinfo{year}{2010}).

\bibitem[{\citenamefont{Oganesyan et~al.}(2001)\citenamefont{Oganesyan,
   Kivelson, and Fradkin}}]{oganesyan2001}
\bibinfo{author}{\bibfnamefont{V.}~\bibnamefont{Oganesyan}},
   \bibinfo{author}{\bibfnamefont{S.~A.} \bibnamefont{Kivelson}},
   \bibnamefont{and} 
\bibinfo{author}{\bibfnamefont{E.}~\bibnamefont{Fradkin}},
   \bibinfo{journal}{Phys. Rev. B} \textbf{\bibinfo{volume}{64}},
   \bibinfo{pages}{195109} (\bibinfo{year}{2001}).

\bibitem{yamase2004} H. Yamase, Phys. Rev. Lett. 93, 266404 (2004)

\bibitem[{\citenamefont{Kao and Kee}(2007)}]{kao2007}
\bibinfo{author}{\bibfnamefont{Y.-J.} \bibnamefont{Kao}} \bibnamefont{and}
   \bibinfo{author}{\bibfnamefont{H.-Y.} \bibnamefont{Kee}},
   \bibinfo{journal}{Phys. Rev. B} \textbf{\bibinfo{volume}{76}},
   \bibinfo{pages}{045106} (\bibinfo{year}{2007}).

\bibitem[{\citenamefont{Lee and Wu}(2009)}]{lee2009nematic}
\bibinfo{author}{\bibfnamefont{W.-C.} \bibnamefont{Lee}} \bibnamefont{and}
   \bibinfo{author}{\bibfnamefont{C.}~\bibnamefont{Wu}}, 
\bibinfo{journal}{Phys.
   Rev. B} \textbf{\bibinfo{volume}{80}}, \bibinfo{pages}{104438}
   (\bibinfo{year}{2009}).

\bibitem[{\citenamefont{Ying et~al.}(2011)\citenamefont{Ying, Wang, Wu, 
Xiang,
   Liu, Yan, Wang, Zhang, Ye, Cheng et~al.}}]{ying2011}
\bibinfo{author}{\bibfnamefont{J.~J.} \bibnamefont{Ying}},
   \bibinfo{author}{\bibfnamefont{X.~F.} \bibnamefont{Wang}},
   \bibinfo{author}{\bibfnamefont{T.}~\bibnamefont{Wu}},
   \bibinfo{author}{\bibfnamefont{Z.~J.} \bibnamefont{Xiang}},
   \bibinfo{author}{\bibfnamefont{R.~H.} \bibnamefont{Liu}},
   \bibinfo{author}{\bibfnamefont{Y.~J.} \bibnamefont{Yan}},
   \bibinfo{author}{\bibfnamefont{A.~F.} \bibnamefont{Wang}},
   \bibinfo{author}{\bibfnamefont{M.}~\bibnamefont{Zhang}},
   \bibinfo{author}{\bibfnamefont{G.~J.} \bibnamefont{Ye}},
   \bibinfo{author}{\bibfnamefont{P.}~\bibnamefont{Cheng}},
   \bibnamefont{et~al.}, \bibinfo{journal}{Phys. Rev. Lett.}
   \textbf{\bibinfo{volume}{107}}, \bibinfo{pages}{067001}
   (\bibinfo{year}{2011}).

\bibitem{preempty} R. M. Fernandes, A. V. Chubukov, J. Knolle, I. Eremin, and J. Schmalian, Phys. Rev. B {\bf 85}, 024534 (2012).

\bibitem{leewc2009} Wei-Cheng Lee and Congjun Wu, Phys. Rev. B {\bf 80}, 104438 (2009).

\bibitem{raghu2009} S. Raghu, A. Paramekanti, E A. Kim, R. A. Borzi, S. A. Grigera, A. P. Mackenzie, and S. A. Kivelson, Phys. Rev. B {\bf 79}, 214402 (2009).

\bibitem{coleman} P. Coleman, private communications.

\bibitem{onari2010} S. Onari and H. Kontani, Phys. Rev. B 85, 134507 (2012). Hiroshi Kontani, Tetsuro Saito, Seiichiro Onari, Phys. Rev. B 84, 024528 (2011).

\bibitem{note} As mentioned in Section IV, the non-Fermi liquid behaviour will eventually disappear when the temperature is low enough, and the point contact spectroscopy described in Ref. [\onlinecite{arham2011,arham2012}] indeed revealed more complicated structures for very low temperature. These complicated structures might result from an interplay between the orbital order and also stripe-like AFM order, which can not be captured by the present theory.

\end{thebibliography}
\end{document}